\documentclass[12pt]{article}
\usepackage{graphicx}

\newcommand\pubnumber{Author's note number}
\newcommand\pubdate{\today}

\def\Title#1{\begin{center} {\Large #1 } \end{center}}
\def\Author#1{\begin{center}{ \sc #1} \end{center}}
\def\Address#1{\begin{center}{ \it #1} \end{center}}

\newcommand\pubblock{\rightline{\begin{tabular}{l} \pubnumber\\
         \pubdate  \end{tabular}}}
\newenvironment{Abstract}{\begin{quotation}  }{\end{quotation}}
\newenvironment{Presented}{\begin{quotation} \begin{center} 
             PRESENTED AT\end{center}\bigskip 
      \begin{center}\begin{large}}{\end{large}\end{center} \end{quotation}}
\begin{document}
\begin{titlepage}
 \pubblock
\vfill
\Title{Exploration of hadronization through heavy flavor production at the future Electron-Ion Collider}
\vfill
\Author{Xuan Li}
\Address{Physics Division, Los Alamos National Laboratory}
\vfill
\begin{Abstract}
The future Electron-Ion Collider (EIC), which is expected to start construction at Brookhaven National Laboratory in 2025, will utilize high-luminosity high-energy electron+proton and electron+nucleus collisions to explore several fundamental questions in the high energy and nuclear physics fields. Exploring how matter is formed from quarks and gluons, which is referred to as the hadronization process, is one of the EIC science objectives. The EIC project detector design led by the ePIC collaboration can realize a series of high precision heavy flavor hadron and jet measurements. Heavy flavor jet substructure and heavy flavor hadrons inside jets, which can provide direct information about the heavy quark hadronization process, have been studied in simulation for electron+proton and electron+nucleus collisions at EIC. The associated physics projections and comparison with latest theoretical calculations will be presented.
\end{Abstract}
\vfill
\begin{Presented}
DIS2023: XXX International Workshop on Deep-Inelastic Scattering and
Related Subjects, \\
Michigan State University, USA, 27-31 March 2023 \\
     \includegraphics[width=9cm]{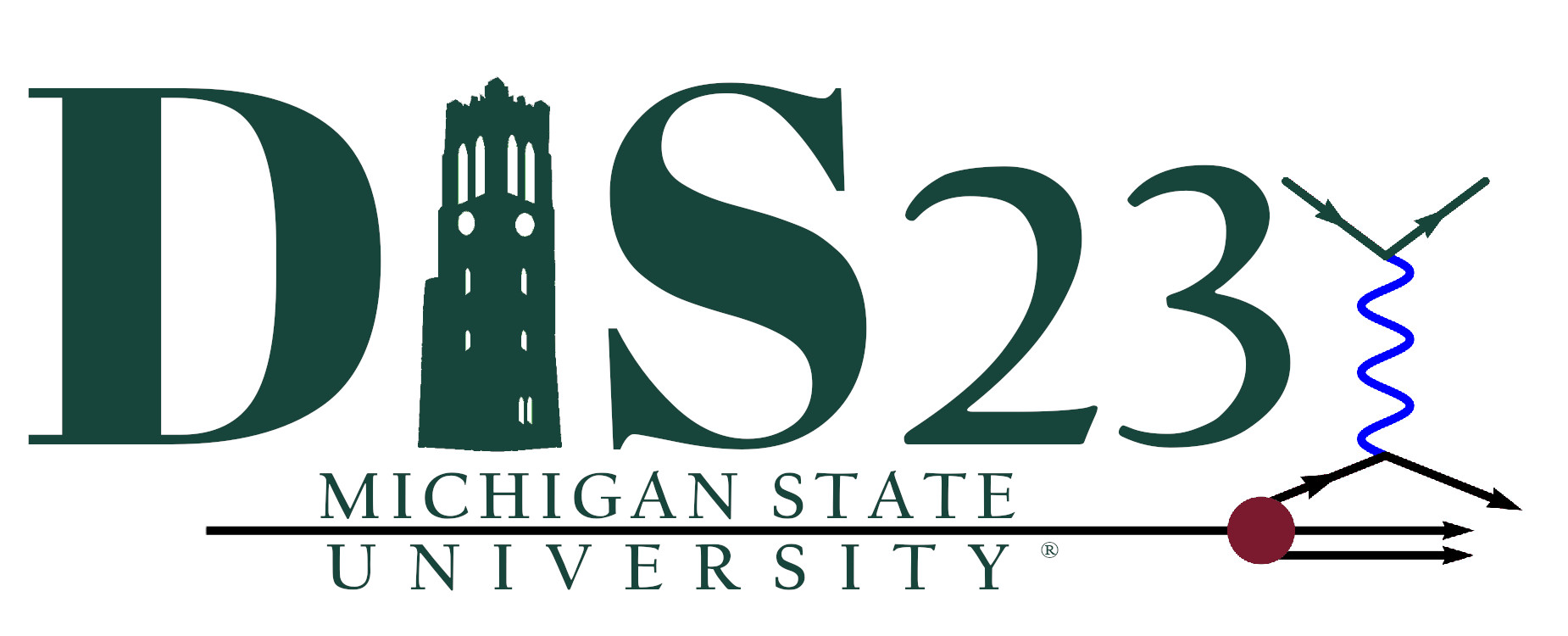}
\end{Presented}
\vfill
\end{titlepage}

\section{Introduction}
\label{sec:intro}
\begin{figure}[tbh]
\centering
\includegraphics[width=0.8\textwidth]{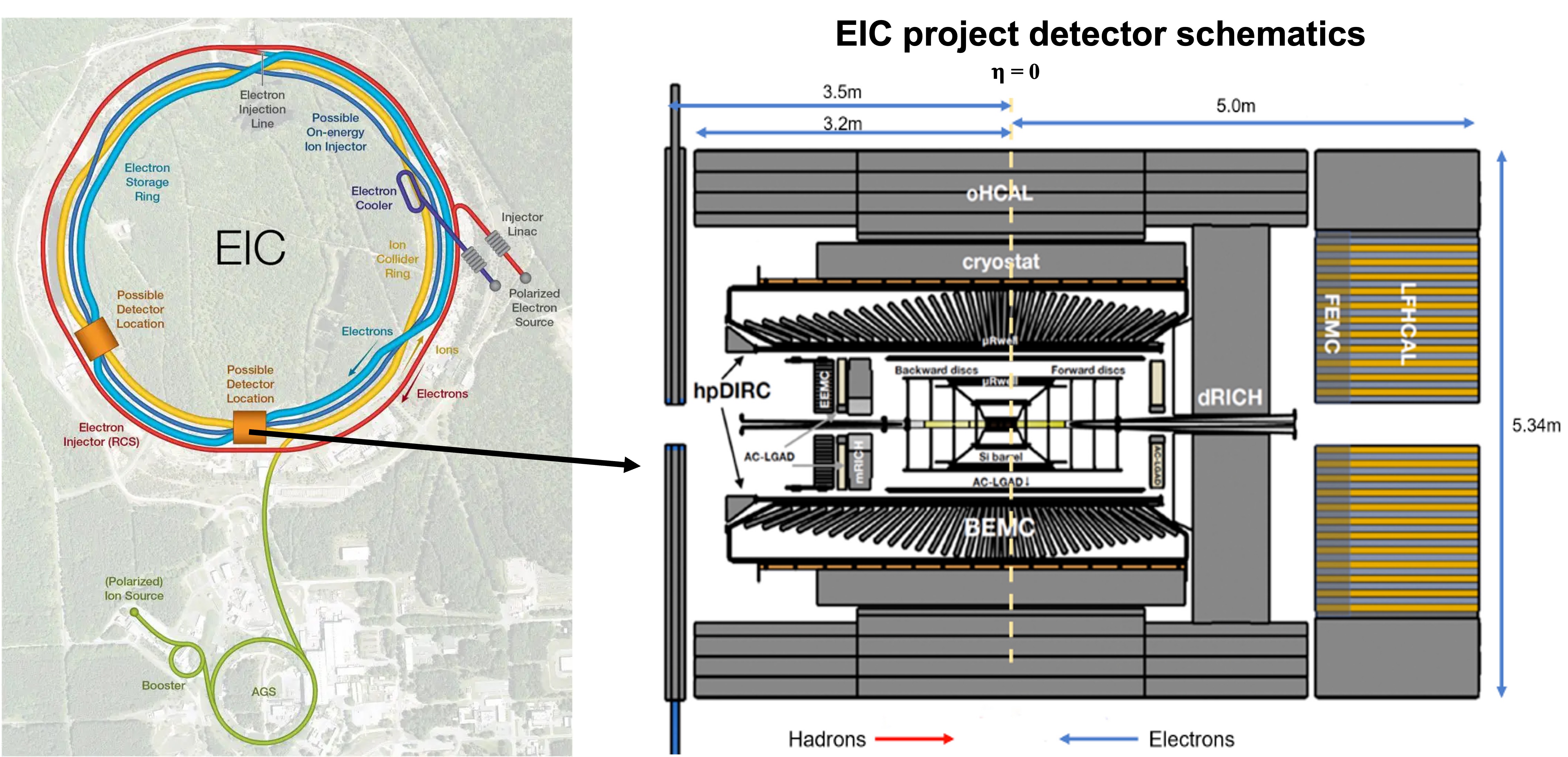}
\caption{Schematics of the EIC accelerator (left) and the current ePIC detector design for the EIC project detector, which is at the 6 o'clock location of the EIC (right). From the inner to the outer region, the ePIC detector consists of high granularity vertex and tracking, particle identification and calorimeter subsystems. A new 1.7~T solenoid magnet will be built for the EIC project detector.}
\label{fig:eic_design}
\end{figure}

The next generation Quantum Chromodynamics (QCD) facility: the Electron-Ion Collider (EIC) is expected to start construction in 2025 at Brookhaven National Laboratory. The future EIC will operate electron+proton ($e+p$) and electron+nucleus ($e+A$) collisions with the instantaneous luminosity at $10^{33-34}$ cm$^{-2}$s$^{-1}$ and the bunch crossing rate at around 10~ns. The electron beam energy varies from 5~GeV to 18~GeV and the proton/nucleus beam energies are at 41~GeV, 100-275~GeV \cite{eic_YR}. The EIC is expected to host two detector experiments at the 6 o'clock and 8 o'clock positions and the detector at the 6 o'clock position is referred to as the project detector (see Figure~\ref{fig:eic_design}). The EIC project detector design optimization is led by the ePIC collaboration. As shown in the right panel of Figure~\ref{fig:eic_design}, the current ePIC detector design consists of high granularity vertex and tracking, particle identification and calorimeter subsystems with the pseudorapidity coverage of $|\eta| \le 3.5$.

One of the EIC science objectives is to explore how quarks and gluons form visible matter inside vacuum or a nuclear medium, which is referred to as the hadronization process. Unlike light quarks, charm and bottom quarks are produced early in collisions and don't change into other quarks or gluons after they are created. How heavy quarks transport inside the nuclear medium and transform into final states hadrons is expected to be different from light quarks. Due to these features, heavy flavor production is a unique probe to explore the hadronization process especially in the little constrained kinematic region \cite{lanl_hf}. The analysis framework in simulation, which includes the event generation in PYTHIA8 \cite{py8}, detector performance evaluation in GEANT4 \cite{geant4}, beam remnant and Quantum Chromadynamic (QCD) backgrounds, and heavy flavor hadron and jet reconstruction algorithms, has been developed for the EIC heavy flavor physics studies. The physics projection of the heavy flavor jet substructure and heavy flavor hadron inside jet nuclear modification factor with the EIC project detector performance in $e+p$ and $e+Au$ collisions will be presented.

\section{EIC heavy flavor hadron and jet reconstruction in simulation}
\label{sec:reco}

\begin{figure}[ht]
\centering
\includegraphics[width=0.96\textwidth]{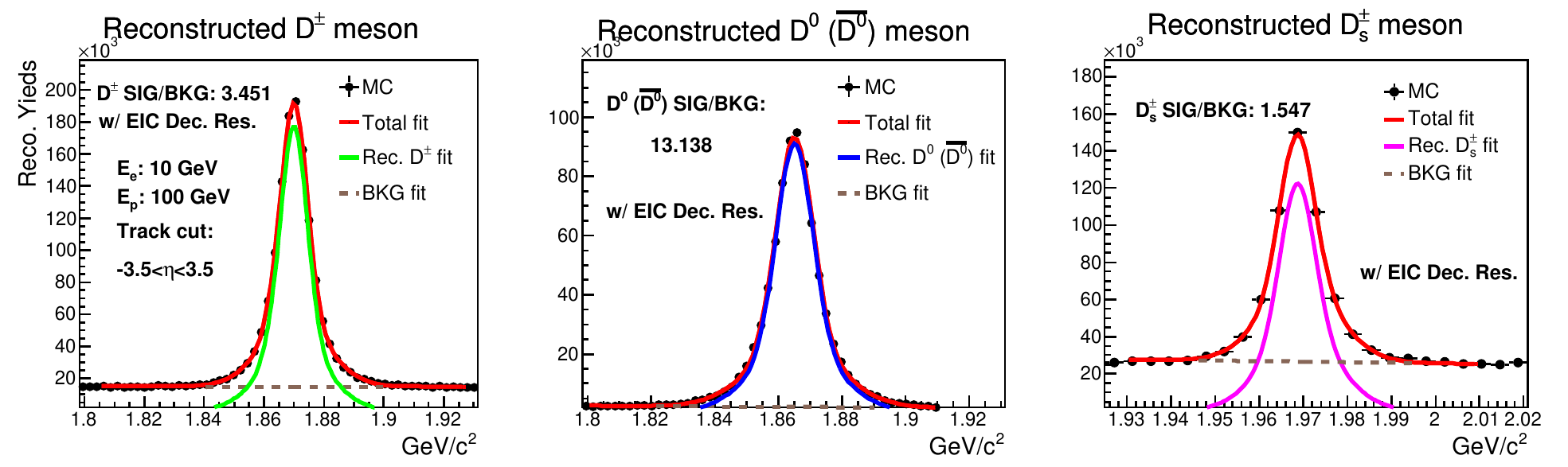}
\includegraphics[width=0.64\textwidth]{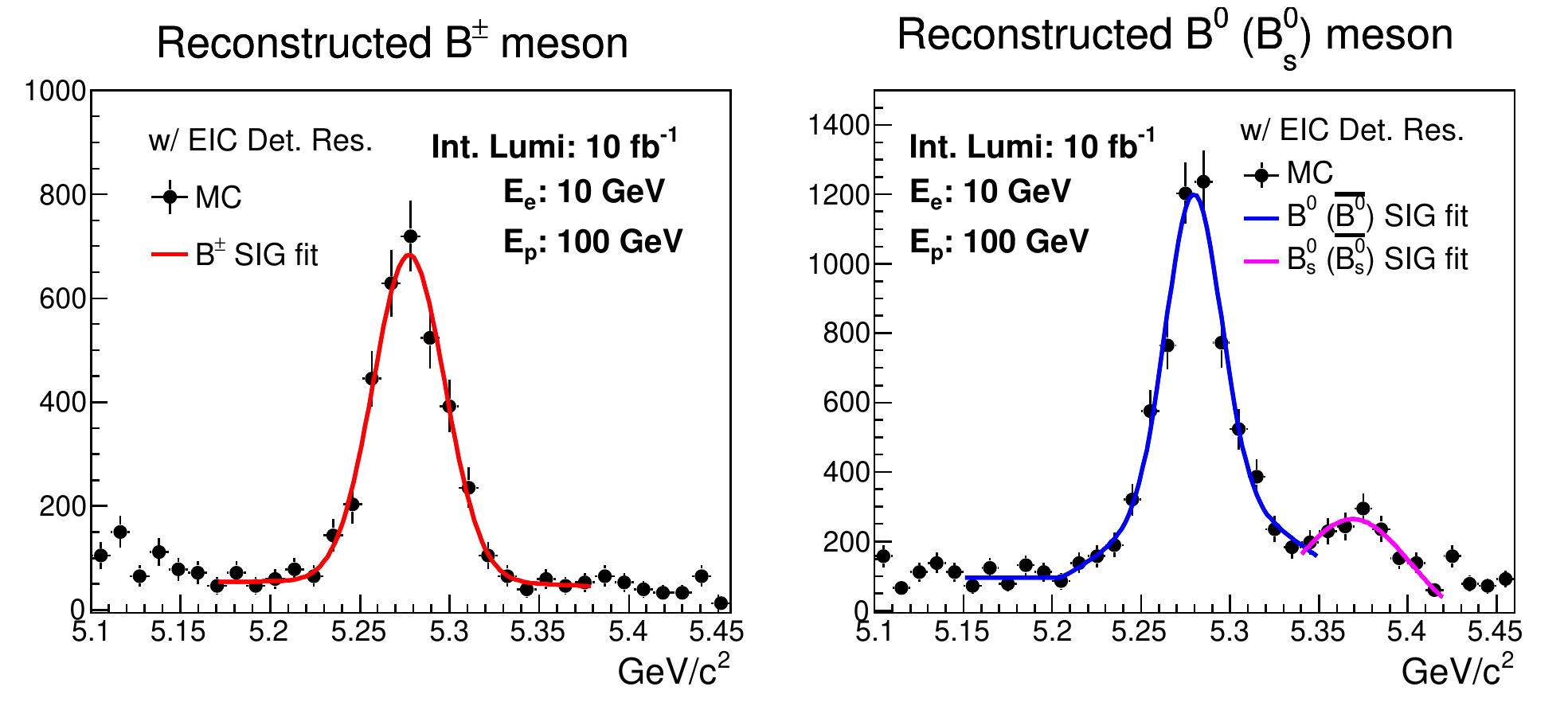}
\includegraphics[width=0.32\textwidth]{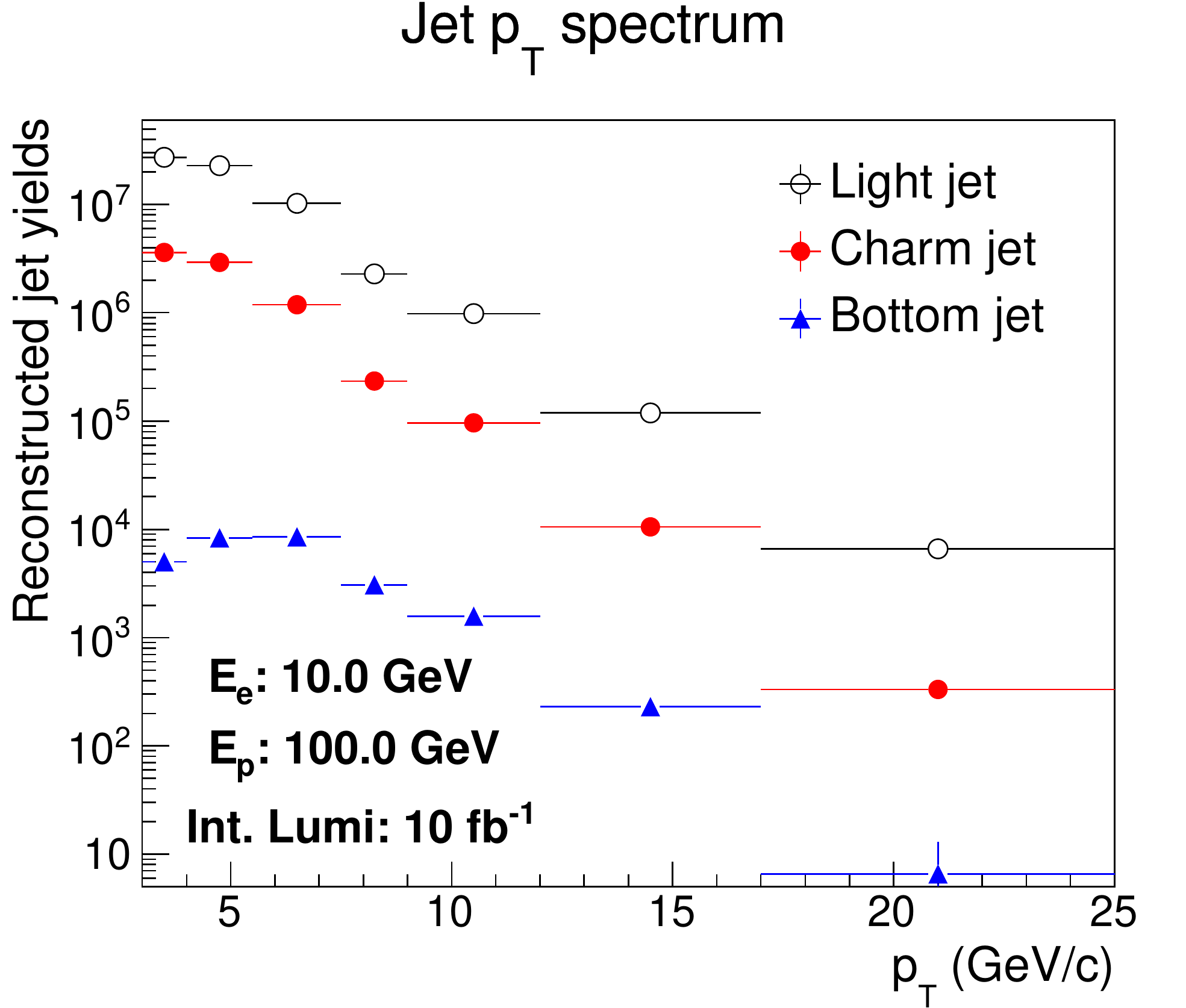}
\caption{The mass spectrum of reconstructed D-mesons ($D^{\pm}$, $D^{0}$ ($\bar{D^{0}}$), $D_{s}^{\pm}$) and B-mesons ($B^{\pm}$, $B^{0}$ ($\bar{B^{0}}$), $B_{s}^{0}$ ($\bar{B_{s}^{0}}$)) and the transverse momentum spectrum of reconstructed jets with different flavors in 63.2 GeV e+p simulation. The reconstructed yields are evaluated with the EIC project detector performance and integrated luminosity of 10~fb$^{-1}$, which is equivalent to around one year EIC operation.}
\label{fig:rec_hf}
\end{figure}

A series of heavy flavor hadron and jet products have been studied with the EIC project detector performance in simulation. As demonstrated in Figure~\ref{fig:rec_hf}, good signal over background ratios have been obtained for reconstructed $D^{\pm}$ ($D^{\pm} \rightarrow K^{\mp}\pi^{\pm}\pi^{\pm}$), $D^{0}$ ($\bar{D^{0}}$) ($D^{0} (\bar{D^{0}}) \rightarrow K^{\mp} \pi^{\pm}$), $D_{s}^{\pm}$ ($D_{s}^{\pm} \rightarrow \phi + \pi^{\pm}$), $B^{\pm}$ ($B^{\pm} \rightarrow J/\psi + K^{\pm}$), $B^{0}$ ($\bar{B^{0}}$) ($B^{0} (\bar{B^{0}}) \rightarrow  J/\psi + K^{\pm} + \pi^{\mp}$), and $B_{s}^{0}$ ($\bar{B_{s}^{0}}$) ($B_{s}^{0} (\bar{B_{s}^{0}}) \rightarrow J/\psi + \phi$) mass spectrums in simulation for 10 GeV electron and 100 GeV proton collisions. As shown in the bottom right panel of Figure~\ref{fig:rec_hf}, reconstructed charm and bottom jets in 63.2~GeV $e+p$ collisions can probe the kinematics of produced charm and bottom quarks in the low transverse momentum region ($p_{T} < 15$ GeV/c) with great precision even with around one year EIC operation. The $e+A$ event generator is under development, therefore the reconstructed heavy flavor hadron and jet yields in $e+A$ collisions are evaluated with the yields of the same product in $e+p$ collisions scaled by the corresponding mass number A.

\section{EIC heavy flavor hadron inside jet physics projection}
\label{sec:hf_in_jet}

\begin{figure}[ht]
\centering
\includegraphics[width=0.96\textwidth]{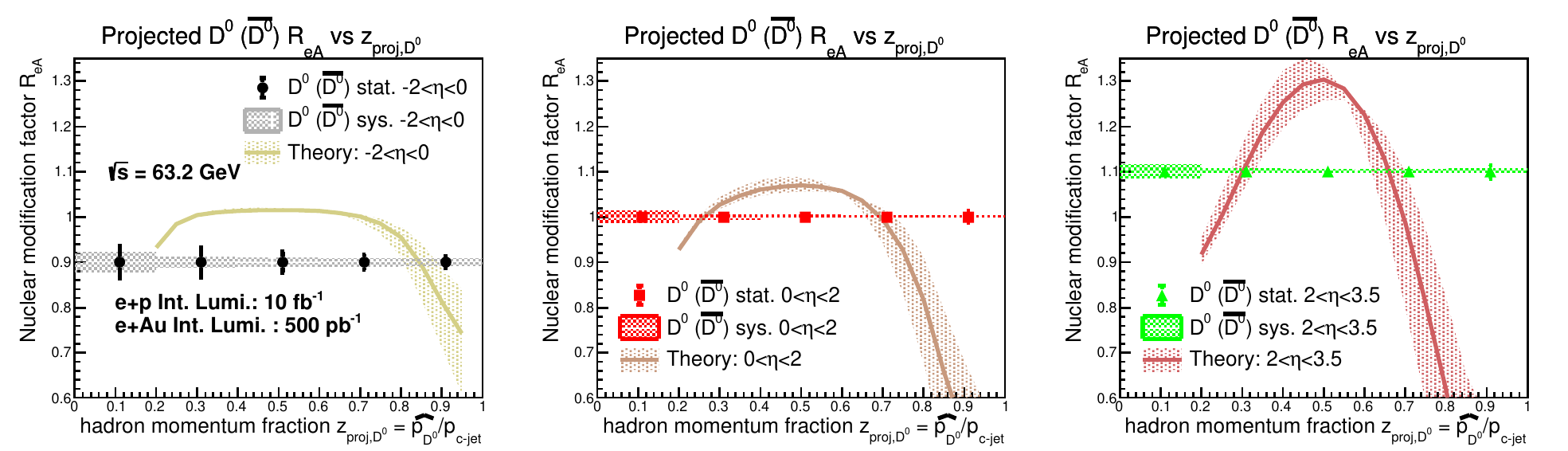}
\caption{The projected accuracy of hadron momentum fraction $z_{proj}$ dependent nuclear modification factor $R_{eAu}$ in 63.2~GeV $e+Au$ collisions with the current conceptual design of the EIC project detector. These projections in the pseudorapidity regions of $-2<\eta<0$ (left), $0<\eta<2$ (middle) and $2<\eta<3.5$ (right) are compared with the Next-to-Leading Order (NLO) perturbative theoretical predictions based on the parton energy loss mechanism.}
\label{fig:d0j_ReA}
\end{figure}

Heavy flavor hadron inside jet production can directly access the heavy quark hadronization process to systematically explore its flavor and kinematic dependence. Better precision of this approach can be achieved in the Deeply Inelastic Scattering (DIS) process as the kinematics of the produced heavy quarks are better constrained in $e+p/A$ collisions than $p+p$/$A+A$ collisions. The hadronization process in $e+p/A$ collisions is expected to be dominated by fragmentation \cite{eic_YR}. One key kinematic variable to access the fragmentation function is the hadron momentum fraction, $z_{proj}$ ($z_{proj} \equiv \vec{p_{h}} \cdot \vec{p_{jet}} / |\vec{p_{jet}}|^{2}$). Figure~\ref{fig:d0j_ReA} shows the hadron momentum fraction $z_{proj}$ dependent projected accuracy of the nuclear modification factor $R_{eAu}$ ($\equiv 1/A  \times \sigma_{e+Au} / \sigma_{e+p}$) of reconstructed $D^{0}$ ($\bar{D^{0}}$) inside charm jets in the pseudorapidity regions of $-2<\eta<0$ (left), $0<\eta<2$ (middle) and $2<\eta<3.5$ (right) in 63.2~GeV e+Au collisions.

Comparison with the latest theoretical predications based on the Next-to-Leading Order (NLO) perturbative parton energy loss calculations \cite{hf_th} indicates that the future EIC charm hadron inside jet measurements will provide significantly better precision in extrapolating the charm quark fragmentation function in medium especially in the high $z_{proj}$ region ($z_{proj} > 0.6$) than existing results. Moreover, the forward D meson inside charm jet measurements at the EIC will help determine the less-known forward charm quark fragmentation function, which is expected to experience a large modification in nuclear medium, with great precision. As more charm and bottom hadrons can be fully reconstructed with the expected performance of the EIC project detector (see Figure~\ref{fig:rec_hf}), this study will be expanded to include more heavy flavor hadron inside jet products to extract the fragmentation functions for different heavy flavor hadron species.

\section{EIC heavy flavor jet substructure study}
\label{sec:hf_jsub}

\begin{figure}[ht]
\centering
\includegraphics[width=0.96\textwidth]{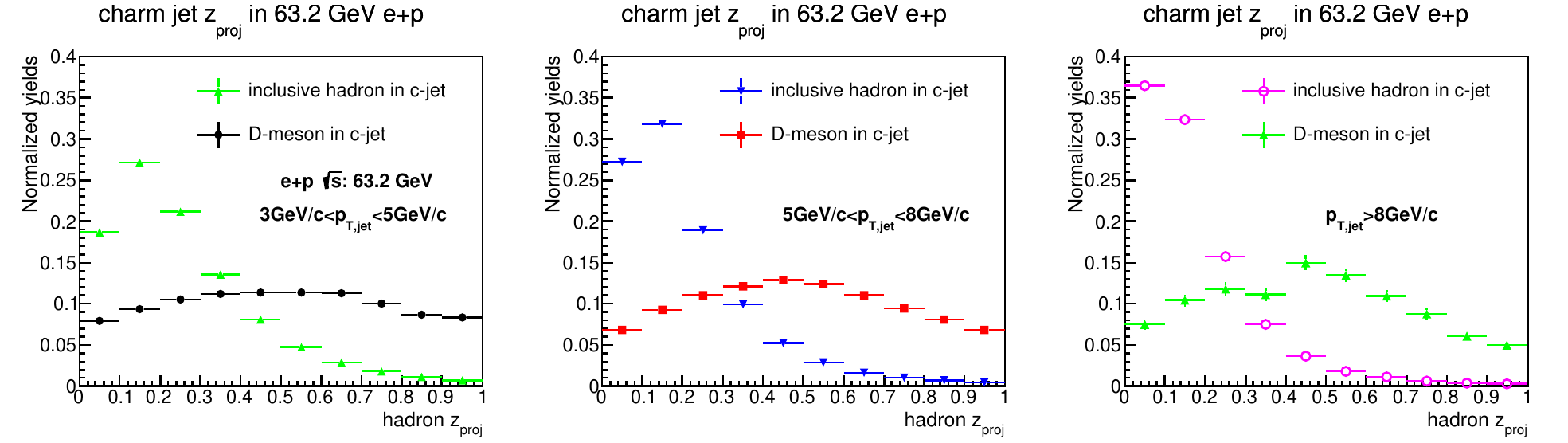}
\caption{Normalized hadron momentum fraction $z_{proj}$ distributions of inclusive hadrons and $D^{0}$ ($\bar{D^{0}}$) mesons inside charm jets in 63.2~GeV $e+p$ simulation. These distributions are evaluated with the EIC project detector performance and divided into three charm jet $p_{T}$ bins: 3~GeV/c $< p_{T, jet} <$ 5~GeV/c (left), 5~GeV/c $< p_{T, jet} <$ 8~GeV/c (middle), and $p_{T, jet} >$ 8~GeV/c (right).}
\label{fig:jet_sub}
\end{figure}

Heavy quarks usually undergo parton shower and energy loss before they transfer into final-state hadrons especially in medium. Separating these final-state effects in heavy flavor cross section measurements will help improve the understanding of the hadronization process. Heavy flavor jet substructure, which is related with the inclusive hadron inside jet production, is one of the ideal probes to extract the heavy quark energy loss and parton shower effects. Figure~\ref{fig:jet_sub} shows the normalized distributions of the hadron momentum fraction $z_{proj}$ carried by inclusive hadrons or $D^{0}$ ($\bar{D^{0}}$) mesons inside charm jets within three jet $p_{T}$ bins: 3~GeV/c $< p_{T, jet} <$ 5~GeV/c (left), 5~GeV/c $< p_{T, jet} <$ 8~GeV/c (middle), and $p_{T, jet} >$ 8~GeV/c (right) in 63.2~GeV $e+p$ collisions. These yields are obtained with the EIC project detector performance in simulation. Enhanced sensitivity to the parton energy loss mechanism will be provided by these low $p_{T}$ heavy flavor jet substructure measurements at the EIC. The shape differences between the $z_{proj}$ distributions of inclusive hadrons inside charm jets and those of $D^{0}$ ($\bar{D^{0}}$) mesons inside charm jets are consistent with the flavor dependent parton showering picture. Further studies of the heavy flavor jet substructure will be expanded to other observables such as the heavy flavor jet angularity \cite{lanl_hf1}, and will be performed in different $e+A$ collisions to extract the medium induced parton showering and energy loss effects.

\section{Summary and Outlook}
\label{sec:sum}
The rapidly evolving EIC project will provide an ideal environment to explore the heavy quark hadronization dynamic process in vacuum and different nuclear medium. The current design of the EIC project detector can realize a series of high precision heavy flavor hadron, jet and hadron inside jet measurements. Initial simulation studies have demonstrated that the proposed EIC heavy flavor hadron inside jet and heavy flavor jet substructure measurements can help extracting the heavy quark energy loss and fragmentation processes in the less constrained kinematic regions with significantly better precisions than existing measurements. As the optimization of the EIC project detector technical design is ongoing, further studies will be carried out with the updated detector performance and include more physics observables.


\begin{thebibliography}{30}
\bibitem{eic_YR}
R. Abdul Khalek et al., 
\emph{Science Requirements and Detector Concepts for the Electron-Ion Collider: EIC Yellow Report},
Nucl. Phys. A 1026 (2022) 122447.
\bibitem{lanl_hf}
X. Li et al.,
\emph{A New Heavy Flavor Program for the Future Electron-Ion Collider}
EPJ Web Conf. 235 (2020) 04002
\bibitem{py8}
T. Torbjorn et al.,
\emph{An introduction to PYTHIA 8.2},
Comput. Phys. Commun. 191 (2015) 159-177.
\bibitem{geant4}
S. Agostinelli, et al., 
\emph{Geant4—a simulation toolkit},
Nuclear Instruments and Methods in Physics Research Section A: Accelerators, Spectrometers, Detectors and Associated Equipment \textbf{506} (3) (2003) 250–303.
\bibitem{hf_th}
H. T. Li, Z. L. Liu, I. Vitev,
\emph{Heavy flavor jet production and substructure in electron-nucleus collisions},
Phys. Lett. B 827 (2022) 137007.
\bibitem{lanl_hf1}
X. Li,
\emph{Heavy Flavor and Jet Studies for the Future Electron-Ion Collider to Explore the Hadronization Process},
SciPost Phys. Proc. 8 (2022), 076.
\end{thebibliography}
\end{document}